\documentclass[12pt]{article}
\usepackage{graphicx}
\begin{document}
\begin{center}
{\Large \bf Ultra High Energy Cosmic Rays from Early Decaying Primordial 
Black Holes}
\end{center}

\bigskip

\begin{center}
{\bf Pankaj Jain}$^1$ 
and {\bf Sukanta Panda}$^2$ 

\bigskip

$^1$Physics Department,\\ Indian Institute of Technology,\\
Kanpur, 208016 India \\

\medskip
$^2$Harish-Chandra Research Institute,\\ Allahabad, India
\end{center}

\begin{abstract}
Origin of ultra high energy cosmic rays is an unsolved problem.  
Several proposals such as Z-burst, decay of super massive matter, susy 
particles as a primary, neutrino as a primary in extra dimension models 
exist in the literature which try to address this issue. Many of these 
proposals solve the problem of propagation of cosmic rays over cosmological 
distances by introducing new physics. However these do not explain the 
origin of such high energy cosmic rays. The possible astrophysics sites, 
such as active galactic nuclei, are highly constrained. Here we determine 
whether these cosmic rays originated from the decay of some exotic objects, 
such as primordial black holes (PBHs), 
present in the early universe. In contrast 
to the usual Top Down scenario we do not assume that this exotic object 
necessarily has to decay in our astrophysical neighbourhood since we assume 
a beyond the standard model scenario, where the propagation problem is 
absent. We consider the standard 4-dimension PBHs as well as 
the brane world PBHs. 
We find that in both cases it is unable to produce the observed
ultra high energy cosmic ray flux. 
\end{abstract}

\newpage

\section{Introduction}
The observation of cosmic rays with energies in excess of $10^{20}\ eV$ 
present a major challenge to astro-particle physics. Due to the presence 
of cosmic microwave background radiation it is predicted that cosmic rays 
with energies above $10^{20}\ eV$ will not be observed due to the
GZK cut-off \cite{gzk}. The presence of 
GZK violating events implies new physics unless the source of these events 
lie within our astrophysical neighbourhood. Some interesting possibilities
include topological defects \cite{topo,bs}, primordial black holes 
\cite{barrau} or super heavy 
particles \cite{shdm} decaying within a distance of about 100 Mpc in order to
evade the GZK bound. However most of such topdown scenarios are severely
constrained by existing data. There also exist many 
proposals which solve the cosmological propagation problem by 
introducing new physics. Examples include  
violations of lorentz invariance \cite{liv,bs}, existence of susy 
particles \cite{susy}, existence 
of magnetic monopoles \cite{monopole}, 
Z-burst \cite{zburst,ss}, a strongly interacting neutrino at ultra 
high energies(UHE) \cite{strongnu} etc. Many of these proposals are also 
severely constrained by existing experiments and will be
further tested by future planned experiments. 
The strongly interacting neutrino proposal, for example, will be ruled out
by non-observation of 
UHE neutrinos in experiments by the year 2006
\cite{ss}.

The only possible conventional 
astrophysical sources for UHE neutrinos are Active Galactic Nuclei 
(AGN) and Gamma Ray Bursts (GRB). These can be considered as possible 
sources of ultra high energy cosmic rays (UHECR) only if we assume
that the propagation problem is solved by some new physics proposal.
It is clearly important to consider 
alternate sources of UHECR, even if they are located at cosmological
distances. Here we consider 
Primordial black holes decaying in the early universe. These objects are 
interesting because, depending on their masses,
they can survive till today. UHECRs from 
Primordial black holes(PBH) in our astrophysical
neighbourhood have been studied in Ref. \cite{barrau}.    

In this paper we consider the production of UHE protons and neutrinos 
from PBHs decaying today and also PBHs which decay in early epoch of
the cosmological evolution of the universe. We calculate UHE fluxes
in standard 4D PBHs as well as in 5D braneworld
PBHs. In the next two sections we review the 4D PBHs and 5D 
brane world PBHs. 

\section{Standard 4D Primordial Black Holes}
It is known that black holes would have formed in very early universe 
through density fluctuations \cite{zn}. These fluctuations 
may either be primordial or may be formed spontaneously at any epoch. 
PBHs might have formed at any spontaneous symmetry breaking epoch through 
the collision of bubbles \cite{cs} or through the collapse of cosmic 
strings \cite{cc}. If one assumes the 
production of black hole with mass of the order of horizon mass at 
some time $t$ in the evolution of universe then its mass \cite{carr},  
\begin{equation}
M_{BH}(t)\approx \frac{c^3 t}{G} \approx 10^{15} 
\left(\frac{t}{10^{-23}s}\right) g.
\end{equation}
Such a black hole with mass of order $10^{15}$g would be 
evaporating now. Masses less 
than $10^{15}$g would have evaporated by now. For example, black holes with 
masses between $10^{10}$ g and $10^{13}$ g would have
completed their evaporation between $10^3$ and $10^{12}$ sec.

The emission rate of spin $1/2$ particles from a black hole 
 is given by \cite{ds},
\begin{equation}
{{\rm d^2}N\over {\rm d}t{\rm d}E} = 
{\Gamma_{1/2}(E,T) \over \exp \left(\frac{E}{kT}\right)+1}    
\label{hawking}
\end{equation}
per particle degree of freedom. Here $T$ is the temperature of the
hole and $\Gamma _{1/2}(E,T)$ is the
absorption coefficient. The above spectrum deviates from the 
black body spectrum due to the energy dependence of
$\Gamma _{1/2}(E,T)$.  
In the limit $E/(kT) \gg 1,$ the spectrum approaches that of a black 
body with a temperature \cite{ds} 
\begin{equation}
T \approx 1.06 \times 10^{13} \left[\frac{1g}{M}\right] \ ,
\label{temp}
\end{equation}
where $M$ is the mass of the black hole. 
For relativistic particles $\Gamma_{1/2}(E,T)$ is given 
by \cite{ds, halzen}, 
\begin{equation}
\Gamma_{1/2}(E,T)\approx \frac{27 E^2}{64 \pi^2 (kT)^2} \ .
\label{gamma}
\end{equation}
A black hole loses mass at the rate \cite{ds, halzen}
\begin{equation}
{{\rm d}M \over {\rm d} t}=-{\alpha (M) \over M^2} \ ,
\label{loss}
\end{equation}
where $\alpha (M)$ depends on the degrees of freedom of the
emitted particles and increases with temperature.
For standard model
\begin{equation}
\alpha(M) \approx 10^{26} g^3 s^{-1} 
\end{equation}
above top quark production threshhold.
From eq. \ref{temp} and eq. \ref{loss}, we find \cite{barrau}
\begin{equation}
dt_* = 1.5 \times 10^{-15} \frac{dT_*}{T_*^4} \ ,
\end{equation}
where $t_*=t/(1 sec)$ and $T_*=T/(1\ EeV).$ Particles with 
energies above 1 EeV will be produced instantly when the temperature 
of the black hole is such that $kT \geq 1\ EeV.$ 
Characteristic time for the 
production of EeV energy particles is of the order of $10^{-18}$ sec 
\cite{barrau}.
Similarly characteristic time required to produce particles with planck 
energy $10^9\ EeV$ is of the order of $10^{-43}$ sec. 
Hence duration of the emission of these high energy particles by PBHs is 
unimportant compared to the evolution time of the universe.  
After integrating eq. \ref{loss}, 
 the mass of the black hole, at any time $t$, is approximately given 
by \cite{ds, halzen, hzmw},
\begin{equation}
M(t)\simeq (M_i^3-3\alpha t )^{1/3}=M_*
\left( \left({M_i\over M_*}\right) ^3-{t\over t_0}\right)^{1/3} \ ,
\label{Mt}
\end{equation}
where $M_i, t_0, M_*$ are respectively the initial mass of the PBHs, age of 
the universe and mass of the PBHs evaporating today. 
The Standard Model provides a lower bound on the high
energy value of $\alpha (M)$, above top production threshold. One can 
assume same value of $\alpha (M)$ to hold at higher energies. In next 
section we review the properties of brane world PBHs. 
\section{Braneworld Primordial Black Holes}
Braneworld cosmological models provide an interesting alternative to the 
standard cosmology. In this scenario
PBHs can be formed in very early universe by density perturbation \cite
{gcl}. In the well known RS2 model 
 our universe is a positive tension brane in a 5d bulk
with $s^1/Z_2$ type of compactification.
It is shown in ref. \cite{mk} that in this model
 PBHs can be easily produced in the 
radiation dominated universe through spherical collapse. 

The Friedmann equation in RS2 model is \cite{gcl}
\begin{equation}
H^2 = \frac{8 \pi}{3 M_4^2} \left(\rho + \frac{\rho^2}{2 \lambda}
+\rho_{kk}\right) + \frac{\Lambda_4}{3} - \frac{k}{a^2}
\label{eq:Friedmann}
\end{equation}
where $H$ is the Hubble constant, $\rho,$ $p$ are respectively the energy
density and pressure of the fluid, $a$ is the scale factor on the brane,  
$M_4$ is the effective 4D planck
mass, $\Lambda_4$ the 4D cosmological constant and $\rho_{kk}$ the
dark radiation density. 
The energy conservation equation on the brane leads to
\begin{equation}
\dot{\rho} + 3 H (\rho + p) = 0 \ .
\end{equation}
As usual
$k=-1,0,1$ for open, flat and closed brane universe. 
 From nucleosynthesis constraints one finds that the 
$\rho_{kk}$ term in equation \ref{eq:Friedmann} is negligible.
The energy density and scale factor then becomes \cite{gcl}
\begin{equation}
\rho = \frac{3 M_4^2}{32 \pi} \frac{1}{t(t+t_c)} 
\end{equation}
and
\begin{equation}
a = a_0 \left( \frac{t (t+t_c)}{t_0 (t_0 + t_c)} \right)^{1/4} \ 
\end{equation}
respectively, 
where $t_0$ is any nonzero time, $t_c$ is the transition time,
$t_c = \frac{l}{2}$,
and $l$ is AdS curvature radius. The time $t_c$ separates
nonconventional cosmology from the standard cosmology. For time $t \gg t_c$
we recover standard cosmology. For time $t \ll t_c$, which we call high energy
regime, the temperature-time relation modifies to \cite{gcl}
\begin{equation}
\frac{T}{T_4} = \left(\frac{45}{8 \pi^3}\right)^{1/4} g_c^{-1/4}
\left(\frac{l}{l_4}\right)^{-1/4} \left(\frac{t}{t_4}\right)^{-1/4} \ ,
\end{equation}
where $T_4,$ $t_4,$ and $l_4$ are 4-D
Planck temperature, Planck time and
Planck length respectively.
At transition time $t_c,$ transition temperature is \cite{gcl,gcl1}
\begin{equation}
T_c = 3 \times 10^{18} \left(\frac{l}{l_4}\right)^{-1/2}\ GeV \ .
\end{equation}
Since current experiments probe gravity to a length scale of 0.2 mm
and constrain the size of extra dimension to be $l \leq 10^{31} l_4$,
we find that the minimum value of $T_c$ is $10^3\ GeV.$

\subsection{Evaporation rate of Brane World PBHs}
In ref. \cite{gcl} authors have calculated a mass-lifetime relation
for black holes formed on the brane due to collapse of
matter on the brane. Effect of accretion on the lifetime and mass of brane 
world 
PBHs are studied in ref. \cite{gcl1, maj, maj1}. We do not consider the 
effect of accretion in this paper. For a review on brane world PBH, see 
ref. \cite{maj2}. If the size of the black holes $r_0 \ll l$ 
then their geometry is described by 5D schwarchild black holes.
These black holes will emit hawking radiation into the brane as well
as to the bulk. In this approximation, radius  and temperature
of the black hole are given by \cite{gcl},
\begin{equation}
r_0 = \sqrt{\frac{8}{3 \pi}}
\left(\frac{l}{l_4}\right)^{1/2} \left(\frac{M}{M_4}\right)^{1/2} l_4
\end{equation}
and
\begin{equation}
T_{bh} = \frac{1}{2 \pi r_0} \ 
\end{equation}
respectively. Mass loss rate of these black holes is given by \cite{gcl},
\begin{equation}
\frac{dM}{dt} \approx -\frac{16 \pi}{3} {\tilde g} T^2 \ ,
\label{bhloss}
\end{equation}
where
\begin{equation}
\tilde g = \frac{1}{160} g_{brane} + \frac{9 \zeta(5)}{32 \pi^4} g_{bulk} \ .
\end{equation}
In our case $g_{brane}$ dominates and most of energy goes to the
brane. If we consider only the Standard Model degrees of freedom,
$g_{brane} =100.$
From eq. \ref{bhloss} we can derive the lifetime $t_{evap}$ of a black
hole of initial mass $M.$ Lifetime $t_{evap}$ is
\begin{equation}
t_{evap} \approx {\tilde g}^{-1} \frac{l}{l_4} \left(\frac{M}{M_4}
\right)^2 \ .
\end{equation}
eq. \ref{bhloss} gives a relation between time and temperature of the BH
as
\begin{equation}
dt_* = \frac{.009 {\tilde g}^{-1}}{512 \pi A} \frac{dT_*}{T_*^5} \ ,
\label{time}
\end{equation}
where $t_* = t/1 sec $, $T_* = T/1\ EeV$ and $A = \frac{l}{l_4}.$
The parameter 
$A$ basically determines the length $l$ at which 5D BHs dominate the dynamics
and can atmost take value $10^{31}$ \cite{gcl2}. This
value comes from upper limit on the size of extra dimension
constrained by sub-mm gravity experiments. 
From mass-lifetime relation we get a range of $l$ over
which PBH acts as a 5 dimensional black hole. The minimum value of AdS
radius \cite{gcl} is
\begin{equation}
\l_{min} = {\tilde g}^{1/3} \left( \frac{t_{evap}}{t_4}\right)^{1/3} l_4
\end{equation}
and the maximum value of AdS radius $l_{max} = 10^{31} l_4.$
This implies that the 5-dimensional PBHs lie in the range \cite{gcl}
\begin{equation}
M_{min} =  {\tilde g}^{1/2}\left( \frac{l_{max}}{l_4}\right)^{-1/2}
\left( \frac{t_{evap}}{t_4}\right)^{1/2} M_4
\end{equation}
to
\begin{equation}
M_{max} =  {\tilde g}^{1/3}\left( \frac{t_{evap}}{t_4}\right)^{1/3} M_4 \ .
\end{equation}
Similarly temperature of the PBHs ranges from
\begin{equation}
T_{min} = \sqrt{\frac{3}{32 \pi}} {\tilde g}^{-1/4}\left(
\frac{l_{max}}{l_{min}}\right)^{-1/4}
\left(\frac{t_{evap}}{t_4}\right)^{-1/4} T_4
\end{equation}
to
\begin{equation}
T_{max} = \sqrt{\frac{3}{32 \pi}} {\tilde g}^{-1/3}
\left(\frac{t_{evap}}{t_4}\right)^{-1/3} T_4 \ .
\end{equation}
We have listed a range of mass, size and temperature of brane world PBHs 
corresponding to their evaporation in Table. \ref{bw1}.
\section{Mass distribution of black hole}
The mass distribution functions for 4D PBHs formed by the collapse of density 
perturbations have been derived in ref. \cite{carr}. Here we assume that the
mass function is dominated by a particular value, as given in \cite{GL}.
In this case the distribution of the PBHs present throughout the
evolution of the universe from their time of formation can be expressed as,
\begin{equation}
n(M) = N (1+z)^3 \ .
\end{equation}
where $z$ is the redshift at time $t$. Here we have 
neglected the evaporation of 
PBHs from their formation time to their time of evaporation. 
We assume similar mass function for 5D brane world PBHs.

\section{Flux Calculation for 4D PBHs}
We are interested in the flux of particles, specifically neutrinos
and protons, at ultra high energies from PBHs. The neutrinos may
be emitted directly by the PBH, which we refer to as the direct
flux. Alternatively they may be emitted by hadrons and other particles
in the PBH spectrum. We refer to this as the indirect flux. 

\subsection{Direct Neutrino Flux}

Let $f(E_\nu,T)$ represent the direct neutrino flux at energy
$E_\nu$ given in eq. \ref{hawking}, then the diffuse
flux per unit area today is:
\begin{equation}
\frac{dN_{\nu}}{dE_{\nu 0}}
 =\frac{1}{4\pi} \times 1.5 \times 
10^{-15}\int_{z_{min}}^{z_{max}}\int_{kT_{i*}(1+z)}^{kT_{pl*}} 
\frac{d(k T_*)}{(k T_*)^4} \frac{1}{(1+z)^2} \frac{dn}{dz} f(E_{\nu},T) 
dz \ ,  
\label{hawking1}
\end{equation}
where $E_\nu = E_{\nu 0}(1+z)$, $z$ is the redshift at the time of 
emission, $z_{max}$ corresponds to maximum redshift 
from which particles of energies $100$ EeV can reach us 
and $E_{\nu 0}$ is energy of neutrino 
at redshift $z_{min}$. We take $z_{max} \leq 10^7$, which
corresponds to the redshift at which particles of initial energy $10^{19}\ GeV$ 
will be observed at energy of 100 EeV. 

\begin{figure}[t]
\hbox{\hspace{0cm}
\hbox{\includegraphics[scale=1]{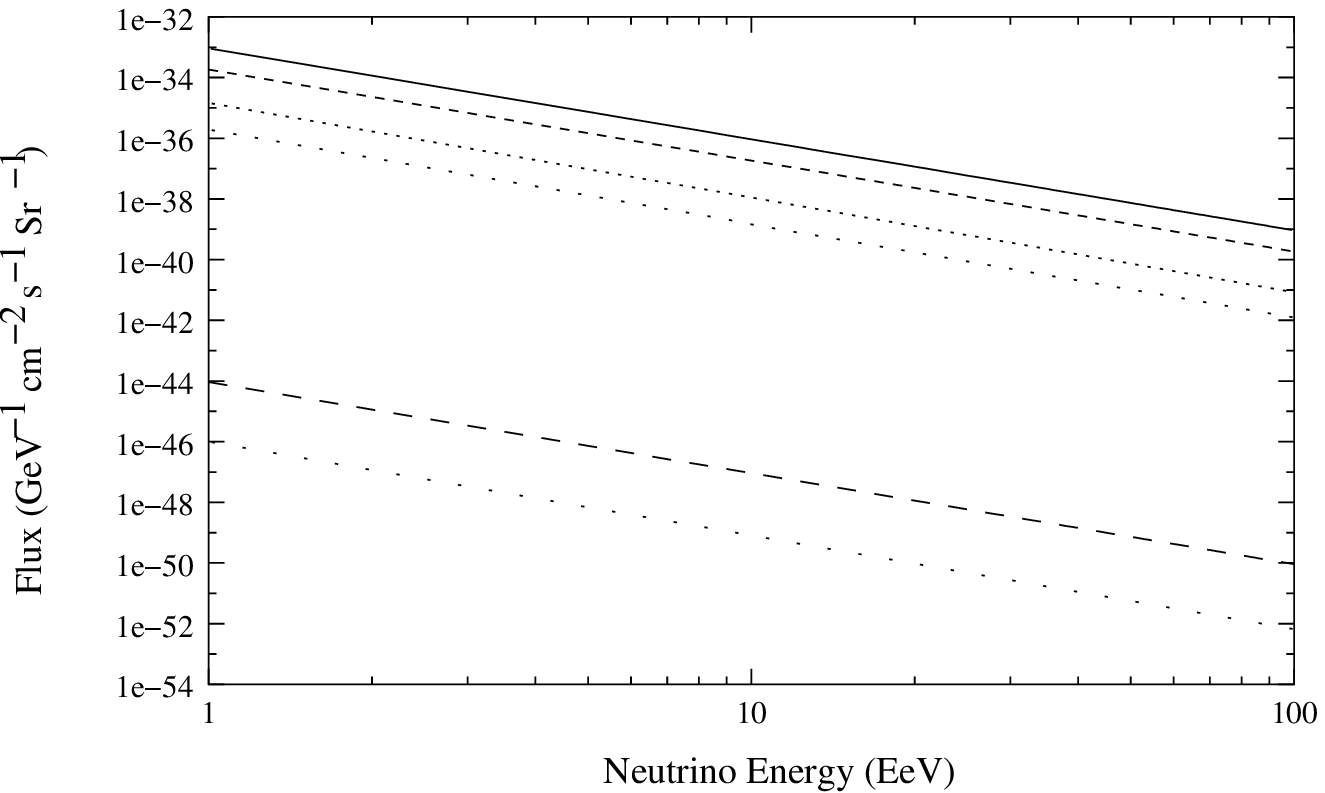}}}
\caption{Direct Neutrino flux today from 4d PBHs evaporating at redshift
$z=0$(solid), $z=1000$(short dashed) and $z=10^6$ (long dashed).
Similarly indirect neutrino flux today from 4d PBHs evaporating at redshift
$z=0$(dottted),
$z=1000$(small spaced dots) and $z=10^6$ (large spaced dots).}
\label{n13}
\end{figure}
\subsection{Indirect Neutrino flux}
The indirect neutrino flux is obtained dominantly from the decay of 
hadrons. The hadrons are formed by the fragmentation of
quarks and gluons. The hadronic flux is generally dominated
by pions which form almost 97 \% of the flux. The remaining 3 \% is mostly
nucleons. The fragmentation function for quarks into hadrons 
is given by
\begin{equation}
x \frac{d N_h}{d x}=\frac{A_h}{x} \exp{-\frac{(\xi-\xi_p)^2}{2 \sigma^2}} 
\end{equation} 
where $\xi_p =Y(\frac{1}{2}+(c_0/Y)^{(1/2)}-c_0/Y)$ and $2\sigma^2 = 
\left(\frac{b Y^3}{36 N_c}
\right)^{1/2}$ with $Y=ln\left(\frac{Q}{\Lambda_{eff}}\right),
b=\frac{11N_c - 2 n_F}{3}$, $c_0=\frac{a^2}{16 b N_c}$ and 
$a=\frac{11N_c}{3}+\frac{2n_f}{3N_c^2}$. The symbols 
$N_c$ and $n_F$ refer to the 
number of colors and number of flavours respectively. 
For $N_c=3$ and $n_F=6$, the parameter $b$ is equal to $7.$
We fix the normalization constant $A_h$ by equating the multiplicity
of corresponding hadrons to their experimental value  
at $Q=\sqrt{s}=91\ GeV$ and $\Lambda_{eff}=200$ MeV.
For pions we find $A_h=4.89.$

In order to calculate indirect neutrino flux we consider the following  
processes. 
\begin{itemize}
\item[1.] the decays of $\mu^+, \mu^-$ and pions.
\item[2.] the fragmentation of quarks into pions and then through the 
following channel $\pi\rightarrow\mu\rightarrow\nu.$
\item[3.] the decays of evaporated W-bosons through the following channel $W
\rightarrow e + \nu$ and $W\rightarrow\mu\rightarrow\nu.$
\end{itemize}

While calculating the indirect neutrino flux we have incorporated the
fragmentation functions by modifying the expression $f(E_{\nu},T).$ 
The final expression for $f(E_{\nu},T)$ in this case is given in the
Appendix of this paper.
\begin{figure}[t]
\hbox{\hspace{0cm}
\hbox{\includegraphics[scale=.95]{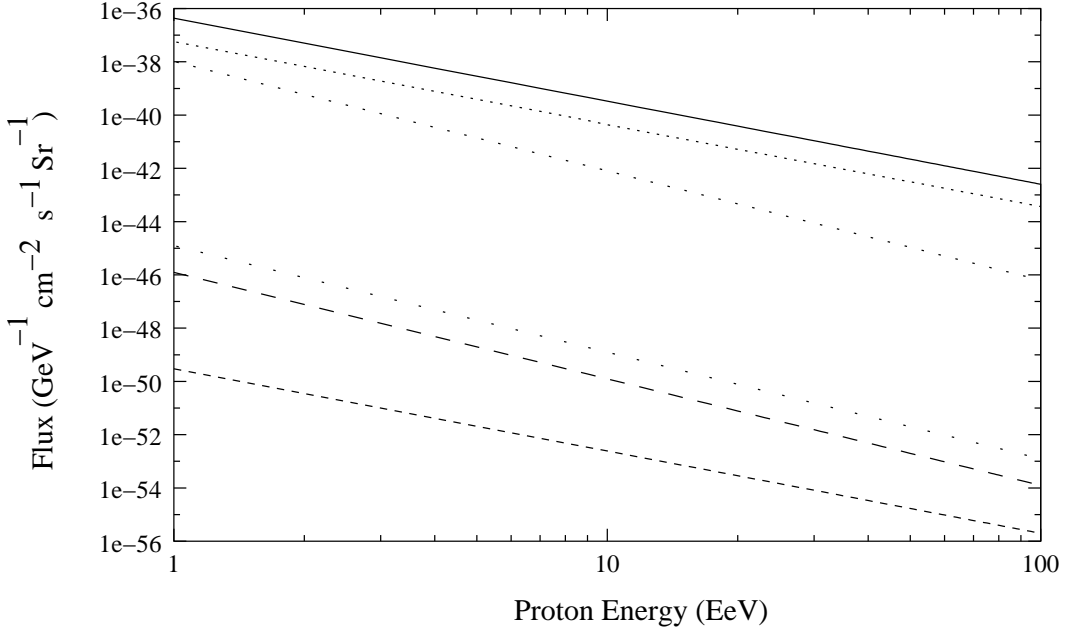}}}
\caption{Proton
flux today from 4d PBHs evaporating at redshift $z=0$(solid), $z=1000$
(dotted) and $z=10^6$ (small spaced dots). Similarly Proton
flux today from 5d BWPBHs evaporating at redshift $z=0$(large spaced dots),
$z=1000$(long dashed) and $z=10^6$ (short dashed).}
\label{p14}
\end{figure}
\subsection{Proton Flux}
The expression for the proton flux is similarly given by
\begin{equation}
\frac{dN_p}{dE_{p0}} = \frac{1}{4\pi} \times 1.5 \times 
10^{-15} \int_{z_{min}}^{z_{max}}{\int_{kT_{i*}(1+z)}^{kT_{*pl}}}  
\int_{E_p}^{\infty} 
\frac{d(kT_*)}{(k T_*)^4}\frac{1}{(1+z)^2} 
\frac{dn}{dz}f(E_q,T) 
\frac{dn_p}{dE_p} dz \ , 
\end{equation}
where $E_p = E_{p 0}(1+z)$, $z$ is the redshift at the time of
emission, $\frac{dn_p}{dx} = .03 \frac{dN_h}{dx},$ $x = \frac{E_p}{E_q}$ 
and $E_{p 0}$ is the energy of proton at redshift $z_{min}.$  
\section{Flux Calculation for 5D brane world PBHs} 
A 5D black hole emits  particles with energy
in the range $(E, E+dE)$ at a rate
\begin{equation}
{{\rm d^2}N\over {\rm d}t{\rm d}E} =
\frac{1}{4 \pi^3} \frac{E^2} {T^2 \exp \left(\frac{E}{kT}\right)+1}
\label{hawk}
\end{equation}
per particle degree of freedom. Here $T$ is the temperature of the
hole.
Let $f(E_{\nu},T)$ represent the total neutrino flux of energy
$E_\nu$ given in eq. \ref{hawk}, then the direct diffuse
neutrino flux per unit area today is:
\begin{equation}
\frac{dN_{\nu}}{dE_{\nu 0}}
 =\frac{.009}{4\pi^4} \frac{{\tilde g}^{-1}}{512 A}
\int_{z_{min}}^{z_{max}}\int_{kT_{i*}(1+z)}^{kT_{pl*}}
\frac{d(k T_*)}{(k T_*)^5} \frac{1}{(1+z)^2} \frac{dn}{dz} f(E_{\nu},T)
dz\,
\label{hawk3}
\end{equation}
where $E_\nu = E_{\nu 0}(1+z)$, $z$ is the redshift at the time of
emission, $z_{max}$ corresponds to max redshift
from which particles of energies $100\ EeV$ can reach us and $z_{min}$
corresponds to minimum redshift. Similarly we calculate the indirect 
neutrino flux and proton flux from braneworld PBHs by incorpoarting 
fragmentation function in eq. \ref{hawk3}.  
\begin{figure}[t]
\hbox{\hspace{0cm}
\hbox{\includegraphics[scale=1]{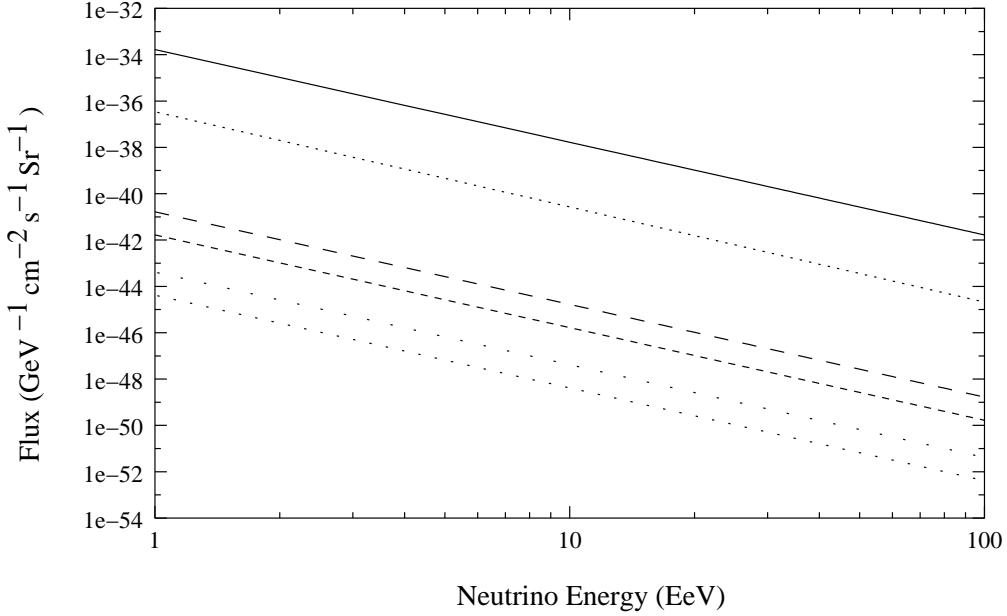}}}
\caption{Direct Neutrino flux today from 5d BWPBHs evaporating at redshift
$z=0$(solid), $z=1000$(dotted) and $z=10^6$ (short dashed).
Similarly indirect neutrino flux today from 4d PBHs evaporating at
redshift $z=0$(dottted),
$z=1000$(long dashed) and $z=10^6$ (large spaced dots).}
\label{dflux}
\end{figure}
\section{Results and Discussion}
The observational constraints on the mass fraction of black holes at 
evaporation is given by the quantity $\alpha_{evap}=
\frac{\rho_{pbh}(M)}{\rho_{rad}}.$
All the constraints with reason are given in table 1 of 
ref. \cite{gl}.  We have taken upper
limits on the PBH densities corresponding to their evaporation time for  
our calculation. The results for the neutrino and proton
flux from 4d primordial 
black holes are given in Fig. \ref{n13} and Fig. \ref{p14} respectively. 
The neutrino flux from the brane world PBH is given in Fig. \ref{dflux}.  
As we see from Figs. \ref{n13} and 
\ref{dflux}, neutrino flux at
energy $10^{20}\ eV$ from PBHs evaporating today
is many orders of magnitude smaller compared to the existing
neutrino flux limit.
For early decaying PBHs it can be noticed that neutrino flux at $10^{20}\
eV$ is even smaller.

 For PBHs of mass $M \approx 5 \times 10^{14}$ g, neutrino flux at 
$10^{20}\ eV$ energy is eight orders of magnitude smaller than the neutrino 
flux limit at that energy. If these PBHs are clustered in galactic halos 
then their density may be somewhat higher.
For PBHs of mass $M \approx 10^{13} g$ and $M \approx 
5 \times 10^{10} g,$ neutrino flux at $10^{20}\ eV$ is very small 
compared to neutrino flux limit  because the number of UHE particles 
emitted by low mass black holes decrease rapidly as their mass decreases. 
Hence, although the constraints on the lower mass PBHs are much weaker, 
the final flux produced by the low mass PBHs turns out to be much smaller.
For $M \approx 10^{13} g$ the 
dominant constraint comes from entropy production at 
nucleosynthesis. While for $M \approx 5 \times 10^{10} g$ the main constraint 
comes from deuterium destruction.  

The astrophysical and cosmological 
constraints on brane world PBHs are obtained in ref. \cite{gcl2,sns,sns1}. 
In ref. \cite{sns,sns1}  constraints on  
brane world PBHs are obtained from high energy diffuse gamma ray 
and from cosmic ray antiproton flux.   
We observe that neutrino fluxes are much smaller compared to 4D 
PBHs even considering  maximum allowed PBH densities at their 
corresponding evaporation era. This is understandable because the temperature
of 5D PBHs is small compared to the 4D PBHs of same mass. UHE neutrino flux 
from brane world PBHs is much smaller compared to 4D PBHs of same mass. 
     
We also calculate proton flux from 4D PBHs and brane world 5D PBHs. 
It turns out that proton flux at $10^{20}\ eV$ from 4D PBHs decaying today is 
roughly ten orders of magnitude smaller than the UHECR flux. Proton flux 
from early decaying 4D PBHs and brane world PBHs are even smaller.

The main uncertainities in our calculation include: (a) lack 
of information about the degrees of 
freedom available at high energies when black hole temperature exceeds 
the energies currently attained in collider experiments and (b) 
extrapolation of the fragmentation function  
to high energies. These unknowns 
at high energy might change the above picture somewhat but is unlikely
to change our results qualitatively.  

To conclude, we find that PBHs, decaying in the early universe, contribute
negligibly to the ultra high energy cosmic ray flux. 
However it may be interesting to repeat our
calculations for other superheavy particles, decaying in the early universe,
which may give a larger contribution.
\begin{table}[h]
\begin{tabular}{|l|l|l|l|l|l|l|r|}
\hline
  $t_{evap}(sec)$ & $10^{17}$ & $10^{12}$ & $10^4$ \\
\hline
  $l_{min}(cm)$ & $6.214 \times 10^{-14}$ &  $1.338 \times 10^{-15}$ & $2.885 
\times 10^{-18}$ \\
\hline
  $l_{max}(cm)$ & .01 & .01 & .01 \\
\hline
  $M_{min}(g)$ & $1.549 \times 10^9$ & $4.898 \times 10^6 g$  & $4.898 \times 10^2$ \\
\hline
  $M_{max}(g)$ & $3.175 \times 10^{14}$ & $1.339 \times 10^{13}$ & $2.885 
\times 10^{10}$ \\
\hline
  $T_{min}$ & $2.3 \times 10^{-23}T_4$ &  $2.45 \times 10^{-22} T_4$
&  $ 2.45 \times 10^{-20} T_4$ \\
\hline
  $T_{max}$ & $1.71 \times 10^{-20} T_4$ &  $4.053 \times 10^{-19} T_4$ & 
$1.88 \times 10^{-16} T_4$ \\
\hline
\end{tabular}
\caption{ Mass, temperature ranges for BW PBHs at three different epochs.}
\label{bw1}
\end{table}
\section{Appendix}
Here we give expressions for $f(E_{\nu},T)$ for different cases.

1. The indirect neutrino flux due to decay of muons and pions may 
be expressed as,

\begin{equation}
f(E_\nu,T)= B \int f(E_\mu,T) \frac{dn_{\nu}(E_\mu,E_\nu)}{dE_\nu} dE_\mu
\ ,
\end{equation}
where $B$ is the number of degrees of freedom.  
 For the case of muon decay, we can express, 
\begin{equation}
\frac{dn_{\nu}(E_\mu,E_\nu)}{dE_\nu}=\frac{2}{\gamma m_\mu} f(x) \,
\end{equation}
where $x=\frac{2 E_{\nu}}{\gamma m_\mu}$,
\begin{equation}
f(x)=2 x^2 (3-2 x)
\end{equation}
for $\nu_\mu$ and
\begin{equation}
f(x)=12 x^2 (1-x)
\end{equation}
for $\nu_e.$ In the present 
case $B= 4$ because we have contributions from $\mu^{+}$
and $\mu^{-}.$
Similar calculation can be done for $\nu_\mu$ spectrum from pion decay.

2. For the fragmentation of quarks$\rightarrow$ pions
$\rightarrow$muons  $\rightarrow$neutrinos, we have 
\begin{equation}
f(E_\nu,T)= B_q\int f(E_q,T) \frac{dn_{\pi}(E_q,E_\pi)}{dE_\pi}
\frac{dn_{\mu}(E_\mu,E_\pi)}{dE_\mu} d\frac{dn_{\nu}(E_\mu,E_\nu)}{dE_\nu}
dE_q dE_\pi dE_\mu \,
\end{equation}
where $\frac{dn_{\nu}}{dE_\nu}$ is the
neutrino spectrum in a $\mu $ decay and $\frac{dn_{\mu}}{dE_\mu}$
is the muon spectrum in a $\pi$ decay.
$B_q$ is degrees of freedom of all six quarks and six antiquarks.
Each quark has two degrees of freedom. Hence $B_q$=24 in this case.
We can express,
\begin{equation}
\frac{dn_{\nu}(E_mu,E_\nu)}{dE_\nu} = \frac{dn_{\nu}(E_\mu,E_\nu)}{dy}
\frac{dy}{dE_\mu}= (g_0(y) - P g_1(y)) \frac{dy}{dE_\mu}
\label{dnudE}
\end{equation}
in the limit $E_\mu \gg m_\mu, m_e,$ where
\begin{equation}
g_0(y) = \frac{5}{3} - 3 y^2 + \frac{4}{3} y^3
\end{equation}
\begin{equation}
g_1(y) = \frac{1}{3} - 3 y^2 + \frac{8}{3} y^3
\end{equation}
for $\nu_\mu$ and
\begin{equation}
g_0(y) = 2 - 6 y^2 + 4 y^3
\end{equation}
\begin{equation}
g_1(y) = -2- 12 y -18 y^2 + 8 y^3
\end{equation}
for $\nu_e.$ In eq. \ref{dnudE}
$P$ is the projection of the muon spin in the muon rest frame along
the direction of the muon velocity in the laboratory frame
\begin{equation}
P = \frac{2 E_\pi r}{E_\mu (1-r)} - \frac{1+r}{1-r} \,
\end{equation}
where $r=\left(\frac{m_\mu}{m_\pi}\right)^2.$
Similarly muon spectrum from pion decay is
\begin{equation}
\frac{dn_{\mu}(E_\pi,E_\mu)}{dE_\mu} = \frac{1}{1-r} \frac{1}{E_\pi}
\end{equation}

3. Similar expressions can be obtained in case of
neutrinos from W-boson decays.

\end{document}